\documentclass{PoS}

\newcommand{\note}[1]{}                    

\newcommand{\mnote}[1]{}                   

\newcommand{\sixth}{\mbox{\small $\frac{1}{6}$}}         
\newcommand{\half}{\mbox{\small $\frac{1}{2}$}}          
\newcommand{\third}{\mbox{\small $\frac{1}{3}$}}         
\newcommand{\eightthird}{\mbox{\small $\frac{8}{3}$}}    
\newcommand{\qed}{\mbox{\tiny $Q\!E\!D$}}                

\def\lsim{\mathrel{\rlap{\lower4pt\hbox{\hskip1pt$\sim$}}
    \raise1pt\hbox{$<$}}}                
\def\gsim{\mathrel{\rlap{\lower4pt\hbox{\hskip1pt$\sim$}}
    \raise1pt\hbox{$>$}}}                

\title{
\vspace*{-1.25cm}
\begin{minipage}{\textwidth}
\begin{flushright}
\texttt{\footnotesize
PoS(Lattice 2012)135 \\%
ADP-12-46/T813       \\%
DESY 12-206          \\%
Edinburgh 2012/22    \\%
Liverpool LTH 965    \\%
}
\end{flushright}
\end{minipage}\\[15pt]
\vspace*{+1.25cm}
       Octet baryon mass splittings from up-down quark mass differences}

\ShortTitle{Isospin breaking $\ldots$} 

\author{\speaker{R. Horsley}$^{\,a}$,
        J. Najjar$^b$,
        Y. Nakamura$^c$,
        D. Pleiter$^{db}$, 
        P.~E.~L. Rakow$^e$,
        G. Schierholz$^f$
        and J.~M. Zanotti$^g$ \\
        \llap{$^a$} School of Physics and Astronomy,
                    University of Edinburgh,
                    Edinburgh EH9 3JZ, UK \\
        \llap{$^b$} Institut f\"ur Theoretische Physik,
                    Universit\"at Regensburg, 93040 Regensburg, Germany \\
        \llap{$^c$} RIKEN Advanced Institute for Computational Science,
                    Kobe, Hyogo 650-0047, Japan \\
        \llap{$^d$} JSC, J\"ulich Research Centre,
                    52425 J\"ulich, Germany \\
        \llap{$^e$} Theoretical Physics Division,
                    Department of Mathematical Sciences,
                    University of Liverpool,
                    Liverpool L69 3BX, UK \\
        \llap{$^f$} Deutsches Elektronen-Synchrotron DESY,
                    22603 Hamburg, Germany \\
        \llap{$^g$} CSSM, School of Chemistry and Physics,
                    University of Adelaide, Adelaide SA 5005, Australia \\
        E-mail: \email{rhorsley@ph.ed.ac.uk} }

\author{QCDSF--UKQCD Collaboration}

\abstract{
   Using an SU(3) flavour symmetry breaking expansion in the quark
   mass, we determine the QCD component of the neutron-proton, Sigma
   and Xi mass splittings of the baryon octet due to up-down (and strange)
   quark mass differences. Provided the average quark mass is kept constant,
   the expansion coefficients in our procedure can be determined from
   computationally cheaper simulations with mass degenerate sea quarks
   and partially quenched valence quarks. Full details and numerical
   results are given in \protect\cite{horsley12a}.}

\FullConference{The 30th International Symposium on Lattice Field Theory 
                - Lattice 2012,\\
		June 24-29, 2012\\
		Cairns, Australia}

\begin{document}


\section{Introduction} 

\begin{figure}[h]
\begin{minipage}{0.40\textwidth}

      \begin{center}
         \includegraphics[width=5.0cm]{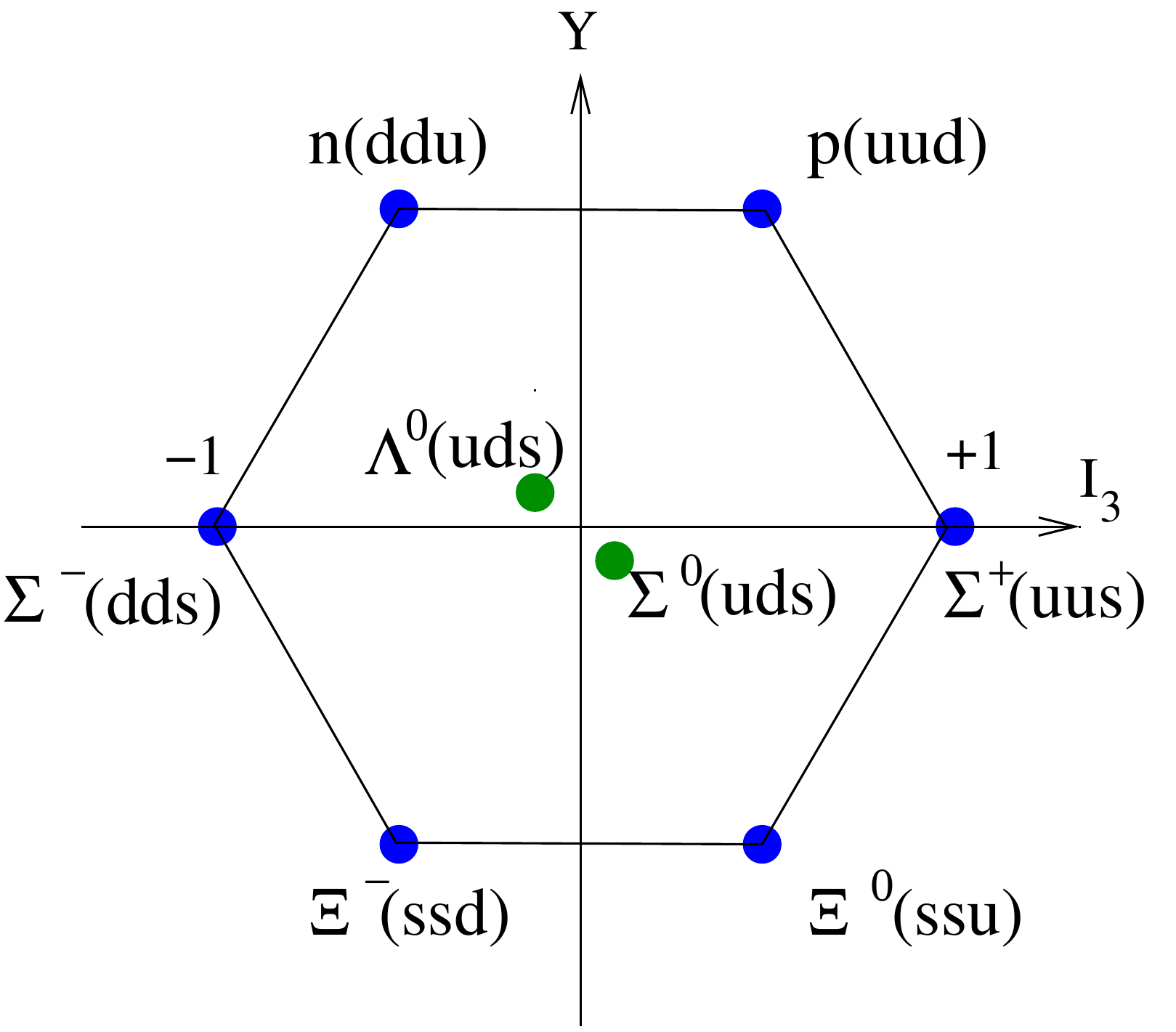}
      \end{center} 

\end{minipage} \hspace*{0.05\textwidth}
\begin{minipage}{0.40\textwidth}

      \begin{center}
         \includegraphics[width=5.0cm]{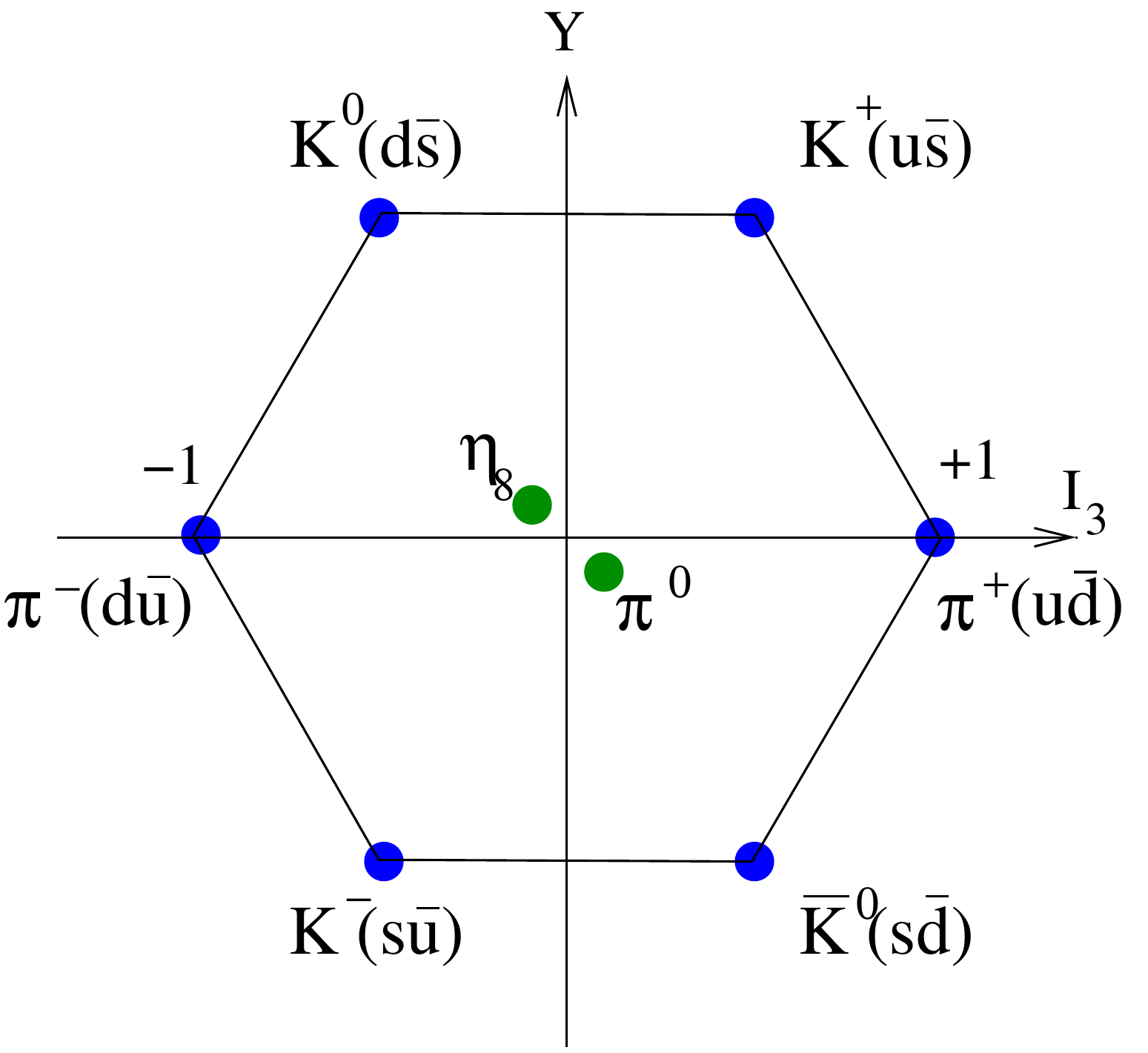}
      \end{center} 

\end{minipage}

\caption{The (lowest) octet baryon states, left panel and
         the octet pseudoscalar states, right panel.}

\label{octet_reps}
\end{figure}

Isospin symmetry was introduced by Heisenberg in the 1930s to explain
non-electromagnetic similarities between the proton and neutron.
Nowadays, of course, this is ascribed to the $u$ and $d$ quarks having
similar mass and the same strong -- or QCD -- interactions.
This $SU(2)$ flavour symmetry is not exact, there are isospin breaking
effects, due to 
\begin{itemize}
   \item the $m_d - m_u$ quark mass difference which is a `pure' QCD effect,
   \item a QED component due to the different quark charges.
\end{itemize}
As both effects are small then we can set
\begin{eqnarray}
   M^{\exp} = M^* + M^{\qed} \,,
\label{qcd_qed_split}
\end{eqnarray}
where we denote by a $^*$ the `pure' QCD component. There is an interplay
between effects: electromagnetic (EM) effects tend to make $p$ heavier
than $n$, but $m_d - m_u$ works in the opposite direction and in fact
dominates as the neutron is heavier than the proton,
$(M_n - M_p)^{\exp} = 1.293333(33)\,\mbox{MeV}$, \cite{nakamura10a}.
\label{pdg_vals}
Including the $s$ quark then the flavour symmetry group becomes $SU(3)$
and the (pseudoscalar) mesons and baryons can be arranged in representations
of this group. In Fig.~\ref{octet_reps} we show the lowest octet baryon and 
pseudoscalar states. States at the center, for example $\Lambda(uds)$,
$\Sigma^0(uds)$ have the same quark content (quantum numbers) but
different wavefunctions and can mix (if isospin is broken)
so we shall only consider states on the `outer' ring here.
As well as $n$ -- $p$ mass splitting we now also have mass
splittings involving the strange quark,
$(M_{\Sigma^-} - M_{\Sigma^+})^{\exp} = 8.079(76)\,\mbox{MeV}$ and
$(M_{\Xi^-} - M_{\Xi^0})^{\exp} = 6.85(21)\,\mbox{MeV}$.
These are all small differences (compared to the masses of the states)
and the experimental precision is way beyond what we achieve here,
but nevertheless we can qualitatively and reasonably quantitatively
describe these splittings, as briefly described in the next section.
For more details see \cite{horsley12a}.


\section{Method}


The QCDSF--UKQCD strategy is to develop an $SU(3)$ flavour symmetry
breaking expansion, \cite{bietenholz11a}, from the flavour symmetric
point down to the physical point. For the baryons on the outer ring
of the octet we have found up to NNLO
\begin{eqnarray}
   M^2(aab) 
      &=& M_0^2 + A_1(2\delta\mu_a+\delta\mu_b) + A_2(\delta\mu_b-\delta\mu_a)
                                                         \nonumber  \\
      & & \phantom{M_0} 
              + \sixth B_0(\delta m_u^2 + \delta m_d^2 + \delta m_s^2)
                                                         \nonumber  \\
      & & \phantom{M_0} + B_1(2\delta\mu_a^2+\delta\mu_b^2)
                        + B_2(\delta\mu_b^2-\delta\mu_a^2) 
                        + B_3(\delta\mu_b-\delta\mu_a)^2
                                                         \nonumber   \\
      & & \phantom{M_0}
          + C_0\delta m_u\delta m_d\delta m_s
                                                         \nonumber   \\
      & & \phantom{M_0}
          + \left[ C_1(2\delta\mu_a + \delta\mu_b) 
                      + C_2(\delta\mu_b - \delta\mu_a)
                \right](\delta m_u^2 + \delta m_d^2 + \delta m_s^2)
                                                         \nonumber   \\
      & & \phantom{M_0}
                 + C_3(\delta\mu_a + \delta\mu_b)^3
                 + C_4(\delta\mu_a + \delta\mu_b)^2(\delta\mu_a - \delta\mu_b)
                                                         \nonumber   \\
      & &  \phantom{M_0}
                 + C_5(\delta\mu_a + \delta\mu_b)(\delta\mu_a - \delta\mu_b)^2
                 + C_6(\delta\mu_a - \delta\mu_b)^3 \,,
\label{SU3_fbreak_octetB}
\end{eqnarray}
while for the pseudoscalar meson octet,
\begin{eqnarray}
   M^2(a\overline{b})
      &=& M^2_{0\pi} + \alpha(\delta\mu_a + \delta\mu_b)
                                                         \nonumber   \\
      & & \phantom{M^2_{0\pi}}
            + \beta_0\sixth(\delta m_u^2 + \delta m_d^2 + \delta m_s^2)
            + \beta_1(\delta\mu_a^2 + \delta\mu_b^2)
            + \beta_2(\delta\mu_a - \delta\mu_b)^2
                                                         \nonumber   \\
      & & \phantom{M^2_{0\pi}}
            + \gamma_0\delta m_u\delta m_d\delta m_s
            + \gamma_1(\delta\mu_a + \delta\mu_b)
                       (\delta m_u^2 + \delta m_d^2 + \delta m_s^2)
                                                         \nonumber   \\
      & & \phantom{M^2_{0\pi}}
           + \gamma_2(\delta\mu_a + \delta\mu_b)^3
           + \gamma_3(\delta\mu_a + \delta\mu_b)
                              (\delta\mu_a - \delta\mu_b)^2 \,.
\label{SU3_fbreak_octetM}
\end{eqnarray}
We have defined for the sea quarks $\delta m_q  = m_q - \overline{m}$
with $\overline{m} = \third(m_u + m_d + m_s)$, where $q \in \{a, b, \ldots\}$
so at the $SU(3)$ flavour symmetric point $\delta m_q = 0$. From this
definition this means that $\delta m_s = -2\delta m_l$. All the expansion
coefficients are functions of $\overline{m}$ only. For the baryon or
meson valence quarks we allow partially quenching, PQ, and set
$\delta\mu_q = \mu_q - \overline{m}$ (i.\ e.\ valence quark masses
$\mu_q \not=$ sea quark masses $m_q$). Of course on the
unitary line when the sea and valence quark masses are the same then
$\delta\mu_q \to \delta m_q$.  The quarks $q = a$, $b$, $\ldots$ are
from $\{u, d, s\}$, so for example $M(uud) = M_p$, $M(dds) = M_{\Sigma^-}$.
We shall also need pseudoscalar mass results and the corresponding
$SU(3)$ flavour breaking expansion to determine the physical point:
$\delta m_d^*$, $\delta m_u^*$ and $\delta m_s^*$ (where a $^*$ denotes
the physical point).

On the unitary line, singlet quantities have the property that
the leading $O(\delta m_q)$ term vanishes. This allows a relatively
simple definition of the scale, as practically we have shown,
\cite{bietenholz11a}, that these quantities hardly vary in the interval
from the flavour symmetric point down to the physical point.
There are many possibilities for example for octet baryons
(which are all stable under strong interactions), we may consider
the `centre of mass' of the octet
\begin{eqnarray}
   X_N^2 = \sixth( M_p^2 + M_n^2 + M_{\Sigma^+}^2 +  M_{\Sigma^-}^2
                        + M_{\Xi^0}^2 + M_{\Xi^-}^2 ) 
         = M_0^2 + O(\delta m_q^2) \,,
\end{eqnarray}
and similarly for the octet of pseudoscalar mesons
\begin{eqnarray}
   X_{\pi}^2 = \sixth( M_{K^+}^2 + M_{K^0}^2 + M_{\pi^+}^2 
                                + M_{\pi^-}^2 + M_{\overline{K}^0}^2 + M_{K^-}^2)
            = M_{0\pi}^2 + O(\delta m_q^2) \,.
\end{eqnarray}
Using this we can form ratios
$\tilde{M} \equiv M / X_S$ for $S = N, \pi$ with expansion coefficients
$\tilde{A}_i \equiv A_i / M_0^2\,,
\tilde{\alpha} \equiv \alpha/M_{0\pi}^2 \,,\ldots$ for the $SU(3)$ 
flavour breaking expansions.

Note that as the coefficients of the $SU(3)$ flavour breaking expansions
are just functions of $\overline{m}$ alone, so provided $\overline{m}$
remains constant, the coefficients can be determined by $n_f = 2+1$
simulations, when $\delta m_u = \delta m_d \equiv \delta m_l$ rather than
more expensive $n_f = 1+1+1$ simulations. Also an additional advantage is
that computationally cheaper PQ results can help to determine the coefficients.

\begin{figure}[h]

\begin{minipage}{0.45\textwidth}

      \begin{center}
         \includegraphics[width=7.00cm]
            {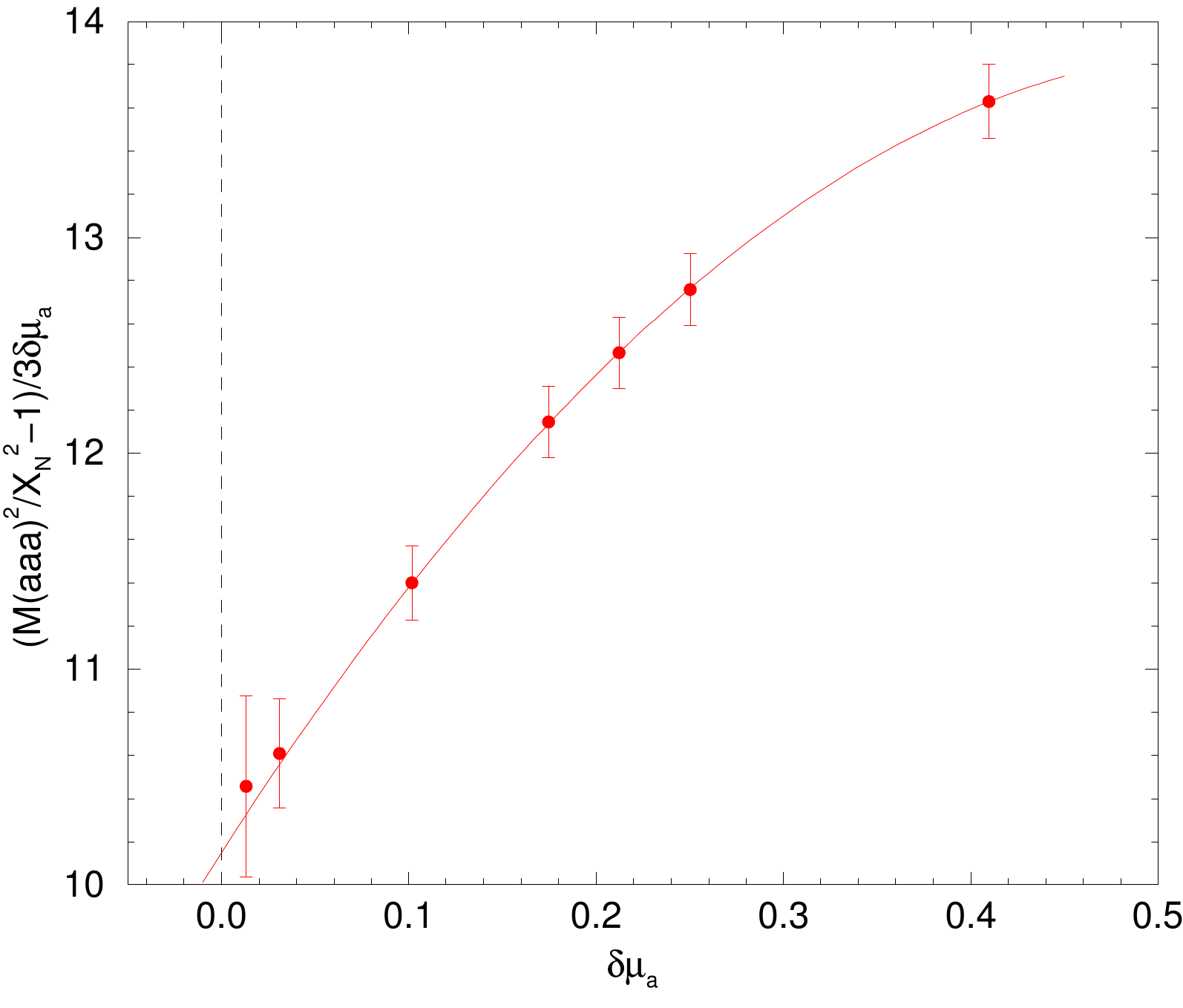}
      \end{center} 

\end{minipage}\hspace*{0.05\textwidth}
\begin{minipage}{0.45\textwidth}

      \begin{center}
         \includegraphics[width=7.00cm]
            {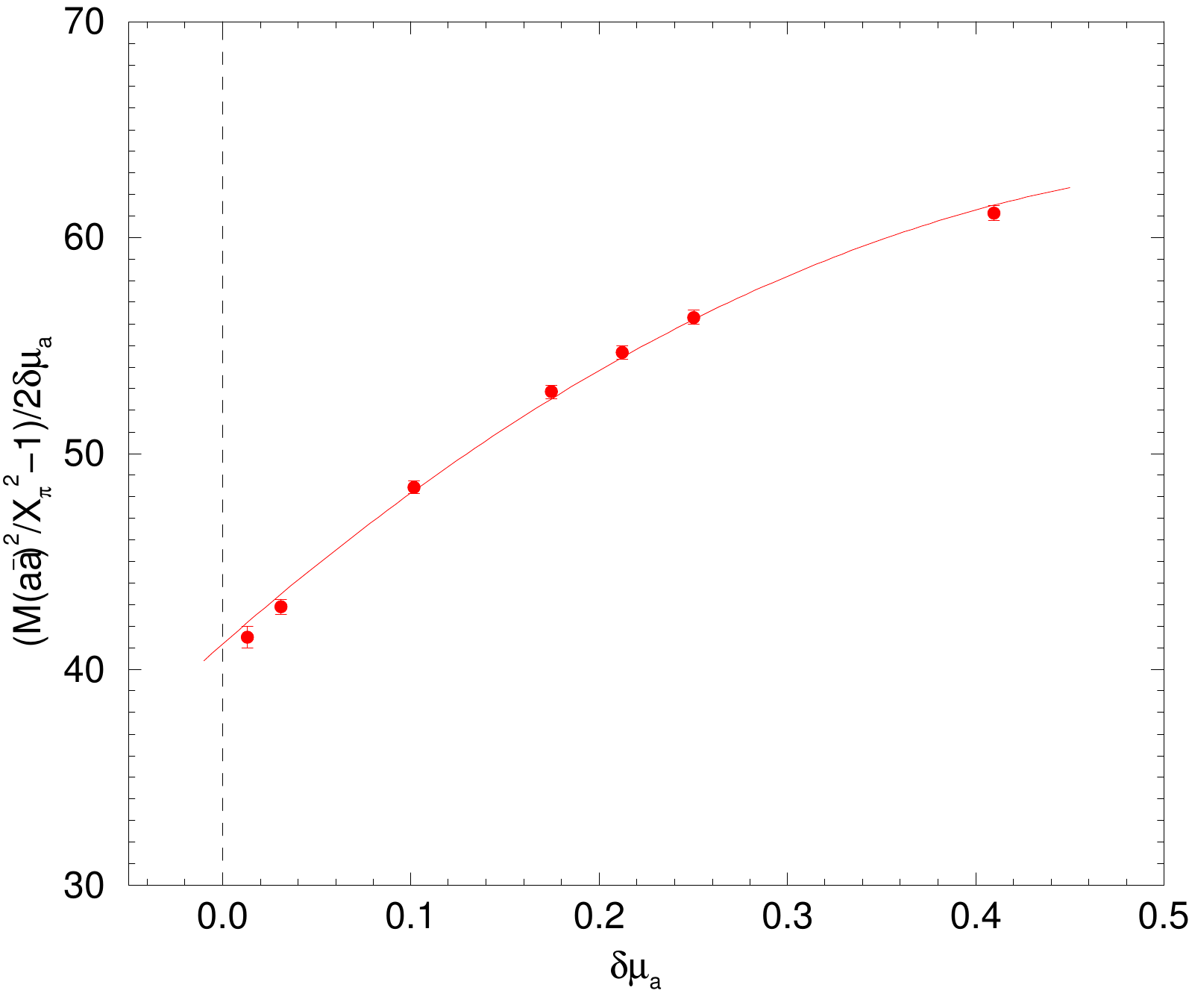}
      \end{center} 

\end{minipage}

\caption{The mass degenerate results, octet baryon $\tilde{M}^2(aaa)$
         left panel and octet pseudoscalar meson $\tilde{M}^2(a\overline{a})$
         right panel (using $32^3\times 64$ sized lattices), both graphs
         versus the PQ quark mass $\delta\mu_a$. Also shown are the 
         functions given in eq.~(\protect\ref{mass_degen_fitfun}).}
\label{PQ_fits}

\end{figure}

\begin{figure}[h]

\begin{minipage}{0.45\textwidth}

      \begin{center}
         \includegraphics[width=7.00cm]
            {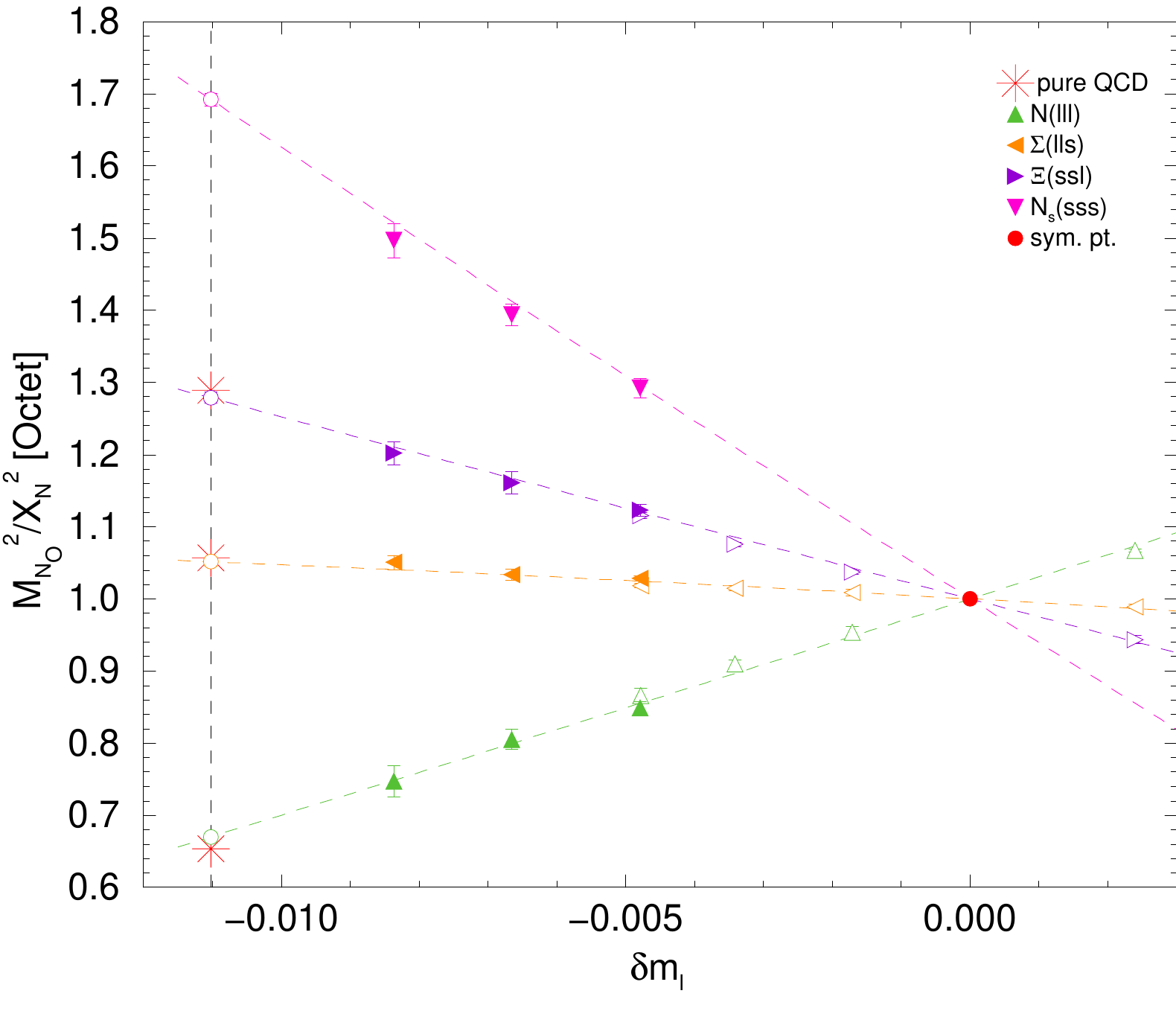}
      \end{center} 

\end{minipage}\hspace*{0.05\textwidth}
\begin{minipage}{0.45\textwidth}

      \begin{center}
         \includegraphics[width=7.00cm]
            {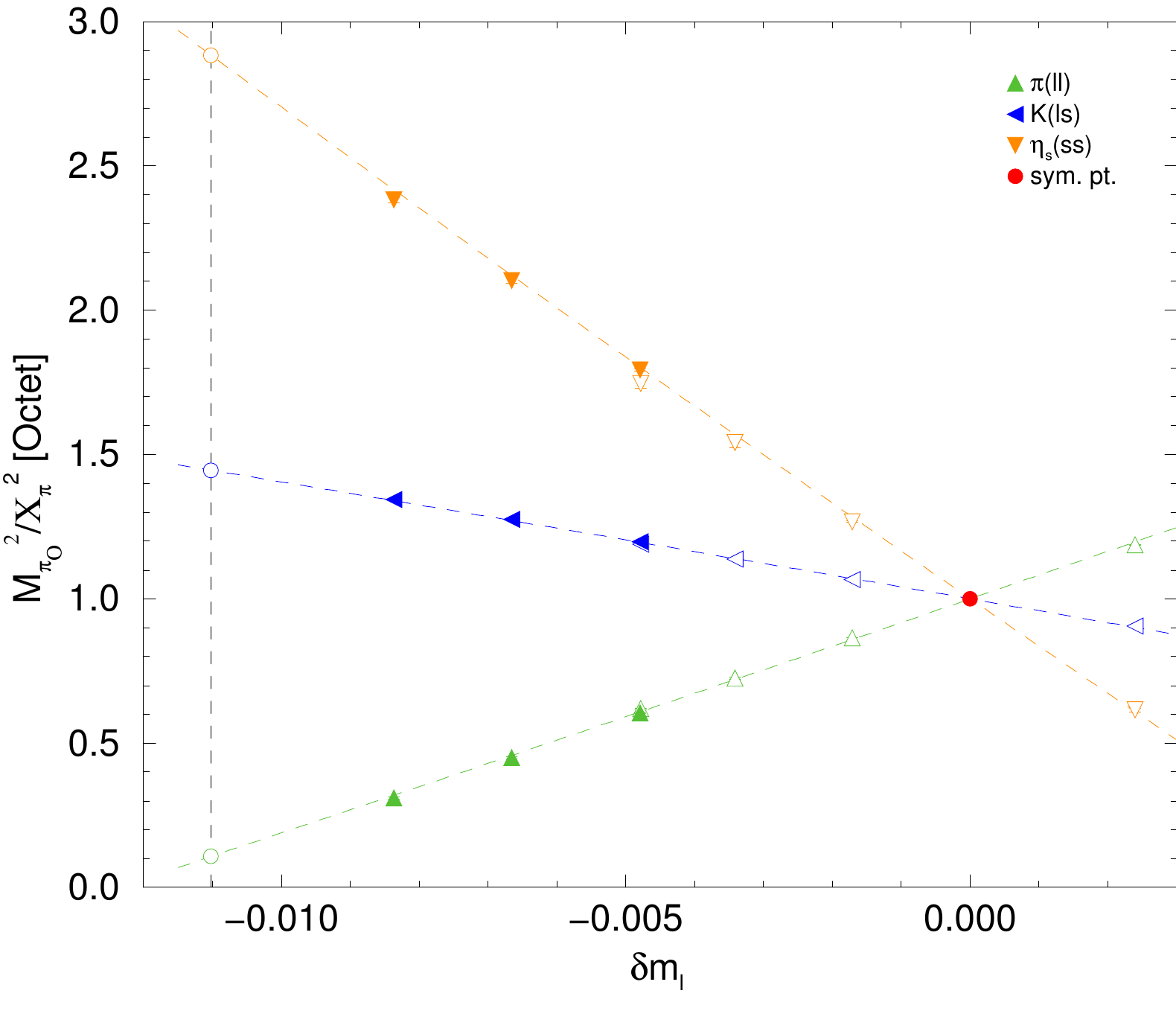}
      \end{center} 

\end{minipage}

\caption{The baryon octet `fan' plot,
         $\tilde{M}^2_{N_O} \equiv {M}^2_{N_O}/X_N^2$
         ($N_O = N$, $\Sigma$, $\Xi$, $N_s$) left panel and
         the pseudoscalar meson octet `fan' plot  
         $\tilde{M}^2_{\pi_O} \equiv {M}^2_{\pi_O}/X_\pi^2$
         ($\pi_O = \pi$, $K$, $\eta_s$) right panel, both graphs
         versus $\delta m_l$.
         The filled symbols represent mass values from $32^3\times 64$
         sized lattices while the opaque symbols are from $24^3\times 48$
         sized lattices (and not used in the fits here).
         The common symmetric point is the filled circle.
         The stars (on the vertical line) are the estimated $n_f = 2+1$
         `pure' QCD physical points. The fits are
         eqs.~(\protect\ref{SU3_fbreak_octetB}) --
         (\protect\ref{SU3_fbreak_octetM}).}

\label{unitary_fits}
\end{figure}

Using $O(a)$-improved clover fermions, \cite{cundy09a}, at $\beta = 5.50$
we have determined the appropriate point on the $SU(3)$ flavour symmetric
line (for the path to the physical point) and then used this point for PQ
determinations of heavier baryon and meson masses (so that $\overline{m}$
obviously remains constant). Fitting these masses (and also including
unitary data at the same constant $\overline{m}$) then allows
determinations of the expansion coefficients. In general these fits are
functions of two quark masses $\delta\mu_a$ and $\delta\mu_b$.
To avoid a $3$--dimensional plot, as an illustration of these fits
in Fig.~\ref{PQ_fits} 
we show $(\tilde{M}^2(aaa)-1)/(3\delta\mu_a)$ (left panel) and
$(\tilde{M}^2(a\overline{a})-1)/(2\delta\mu_a)$ (right panel),
together with the fit functions derived from
eqs.~(\ref{SU3_fbreak_octetB}) -- (\ref{SU3_fbreak_octetM})
by taking completely degenerate quark masses
\begin{eqnarray}
   {\tilde{M}^2(aaa)-1 \over 3\delta\mu_a}
      = \tilde{A}_1 + \tilde{B}_1\delta\mu_a 
                    + \eightthird\tilde{C}_3\delta\mu_a^2 \,, \qquad
   {\tilde{M}^2(a\overline{a})-1 \over 2\delta\mu_a}
      = \tilde{\alpha}_1
                      + \tilde{\beta}_1\delta\mu_a 
                      + 4\tilde{\gamma}_2\delta\mu_a^2 \,.
\label{mass_degen_fitfun}
\end{eqnarray}
As another example, we also consider the unitary results 
(i.e.\ $\delta\mu_q \to \delta m_q$) from the $SU(3)$ flavour
symmetric point down to the physical point. In Fig.~\ref{unitary_fits}
we show the baryon and pseudoscalar octet `fan' plots, where
$M_N \equiv M(lll)$, $M_\Sigma \equiv M(lls)$, $M_\Xi \equiv M(lss)$,
$M_{N_s} \equiv M(sss)$ and $M_\pi \equiv M(l\overline{l})$,
$M_K \equiv M(l\overline{s})$, $M_{\eta_s} \equiv M(s\overline{s})$.
The states $M_{N_s}$, $M_{\eta_s}$ are either not in the octet or are
a PQ state and are not physical, but nevertheless can be used to help
determine the expansion coefficients. The vertical lines are the
$n_f = 2+1$ pure QCD physical point, with the opaque circles being
the determined pure QCD hadron mass ratios for $2+1$ quark flavours.
For comparison, the stars represent the average of the squared masses
of the appropriate particle on the outer ring of the baryon octet,
Fig.~\ref{octet_reps}, i.e.\
$M_N^{*\,2}(lll) = (M_n^{\exp\,2}(ddu) + M_p^{\exp\,2}(uud))/2$,
$M_\Sigma^{*\,2}(lls) 
 = (M_{\Sigma^-}^{\exp\,2}(dds) + M_{\Sigma^+}^{\exp\,2}(uus))/2$,
$M_\Xi^{*\,2}(ssl) = (M_{\Xi^-}^{\exp\,2}(ssd) + M_{\Xi^0}^{\exp\,2}(ssu))/2$.
One immediate observation of Fig.~\ref{unitary_fits} is that there
is hardly any curvature in the data and that the NLO (i.e.\ quadratic
terms) are sufficient.

Finally note that the $x$-scales used in Fig.~\ref{PQ_fits}
and Fig.~\ref{unitary_fits} are very different $|\delta m_l| \sim 0.01
\ll \delta\mu_a \sim 0.5$. (Indeed $\delta\mu_a \sim 0.5$ is roughly
at the charm quark mass.) The ability to use a large range for the
PQ fits enables a much better determination of the fit coefficients
(in particular the NLO terms, which are poorly determined in
the narrow range $|\delta m_l| \sim 0.01$).  

Of course, we are interested in mass differences here. So for the
LO and NLO terms in eq.~(\ref{SU3_fbreak_octetB}) we have along the
unitary line
\begin{eqnarray}
   \tilde{M}_n - \tilde{M}_p 
      &=& \tilde{M}(ddu) - \tilde{M}(uud)
                                                           \nonumber  \\
      &=& (\delta m_d - \delta m_u)
          \left[ \tilde{A}_1^\prime -2\tilde{A}_2^\prime 
                 + (\tilde{B}_1^\prime-2\tilde{B}_2^\prime)
                          (\delta m_d + \delta m_u)
          \right] \,,
                                                           \nonumber  \\
   \tilde{M}_{\Sigma^-} - \tilde{M}_{\Sigma^+} 
      &=& \tilde{M}(dds) - \tilde{M}(uus)
                                                           \nonumber  \\
      &=& (\delta m_d - \delta m_u)
          \left[ 2\tilde{A}_1^\prime - \tilde{A}_2^\prime 
                 + (2\tilde{B}^\prime_1-\tilde{B}_2^\prime + 3\tilde{B}_3^\prime)
                          (\delta m_d + \delta m_u)
          \right] \,,
                                                           \nonumber  \\
   \tilde{M}_{\Xi^-} - \tilde{M}_{\Xi^0} 
      &=& \tilde{M}(ssd) - \tilde{M}(ssu)
                                                           \nonumber  \\
      &=& (\delta m_d - \delta m_u)
          \left[ \tilde{A}_1^\prime + \tilde{A}_2^\prime 
                 + (\tilde{B}_1^\prime + \tilde{B}_2^\prime + 3\tilde{B}_3^\prime)
                          (\delta m_d + \delta m_u)
          \right] \,,
\label{mass_diff}
\end{eqnarray}
where the prime coefficients are simply related to the unprimed ones,
\cite{horsley12a}. Similarly we may invert the meson pseudoscalar expansion,
eq.~(\ref{SU3_fbreak_octetM}), to give
\begin{eqnarray}
   \delta m_d - \delta m_u
      &=& {\tilde{M}_{K^0}^2 - \tilde{M}_{K^+}^2 \over \tilde{\alpha}} \,
          \left( 1 + {2(\tilde{\beta}_1 
                   + 3\tilde{\beta}_2) \over 3\tilde{\alpha}^2}
                       (\half(\tilde{M}_{K^0}^2+\tilde{M}_{K^+}^2)
                                              -\tilde{M}_{\pi^+}^2)
          \right) \,,
                                                           \nonumber \\
   \delta m_d + \delta m_u
      &=& - {2 \over 3\tilde{\alpha}}\,
          \left( \half(\tilde{M}_{K^0}^2+\tilde{M}_{K^+}^2)
                                       -\tilde{M}_{\pi^+}^2
          \right) \,,
\label{q_mass_diff}
\end{eqnarray}
and then substitute in the baryon expansion eq.~(\ref{mass_diff}) to
give the `pure' QCD result.


\section{Results}


Performing this substitution gives the numerical results, \cite{horsley12a},
\begin{eqnarray}
   \tilde{M}_n - \tilde{M}_p
      &=& 0.0789(41)(34) \, \left( \tilde{M}_{K^0}^2-\tilde{M}_{K^+}^2
                          \right) \,
            \left[ 1 +0.0817(92)
                   \left( \half(\tilde{M}_{K^0}^2+\tilde{M}_{K^+}^2)
                                 -\tilde{M}_{\pi^+}^2 \right)
            \right] \,,
                                                           \nonumber \\
   \tilde{M}_{\Sigma^-} - \tilde{M}_{\Sigma^+}
      &=& 0.2243(35)(92) \, \left( \tilde{M}_{K^0}^2-\tilde{M}_{K^+}^2
                        \right) \,
            \left[ 1 + 0.0077(30)
                   \left( \half(\tilde{M}_{K^0}^2+\tilde{M}_{K^+}^2)
                                 -\tilde{M}_{\pi^+}^2 \right)
            \right] \,,
                                                           \nonumber \\
   \tilde{M}_{\Xi^-} - \tilde{M}_{\Xi^0}
      &=& 0.1455(24)(59) \, \left( \tilde{M}_{K^0}^2-\tilde{M}_{K^+}^2
                        \right) \,
            \left[ 1 -0.0324(50)
                   \left( \half(\tilde{M}_{K^0}^2+\tilde{M}_{K^+}^2)
                                 -\tilde{M}_{\pi^+}^2 \right)
            \right] \,,
\label{result_split}
\end{eqnarray}
where $\tilde{M} = M/X_S$, $S = N$, $\pi$. We see that the NLO corrections
are small from $+10\% \sim -5\%$ indicating that the $SU(3)$ flavour
symmetry breaking expansion appears to be a highly convergent series.
(In eq.~(\ref{result_split}) the first error is statistical, the other
is the total systematic error.)

Note that eq.~(\ref{result_split}) is a `pure' QCD result, and is the
main result of this talk. We must now discuss what are the `pure' QCD
values of $M_{K^0}^2$, $M_{K^+}^2$, $M_{\pi^+}^2$. We know that EM
effects are comparable to effects due to $u$ -- $d$ quark mass differences.
Dashen's theorem states that EM effects for charged mesons $K^+$, $\pi^+$
are the same and for neutral mesons $\pi^0$, $K^0$ vanish, Thus
we can write $M_{\pi^+}^{\exp\,2} = M_{\pi^+}^{*\,2} + \mu_\gamma$,
$M_{\pi^0}^{\exp\,2} = M_{\pi^0}^{*\,2} \approx M_{\pi^+}^{*\,2}$,
$M_{K^+}^{\exp\,2} = M_{K^+}^{*\,2} + \mu_\gamma$ and
$M_{K^0}^{\exp\,2} = M_{K^0}^{*\,2}$ where a $^*$ denotes the `pure' QCD
`physical' value, or
$   M_{K^0}^{*\,2} - M_{K^+}^{*\,2} 
      = \left(M_{K^0}^2 - M_{K^+}^2\right)^{\exp}
                 + (1 + \epsilon_\gamma)
                      \left( M_{\pi^+}^2 - M_{\pi^0}^2 \right)^{\exp} \,,$
where violations to Dashen's theorem ($\epsilon_\gamma = 0$) are given by
a non-zero $\epsilon_\gamma$. We shall regard $\epsilon_\gamma$ here as a
possible further systematic error, a typical value for it being
$\epsilon_\gamma =0.7$, \cite{colangelo10a}, giving about a $17\%$
additional systematic error.

\begin{figure}[h]

\begin{minipage}{0.45\textwidth}

      \begin{center}
         \includegraphics[width=5.00cm]
            {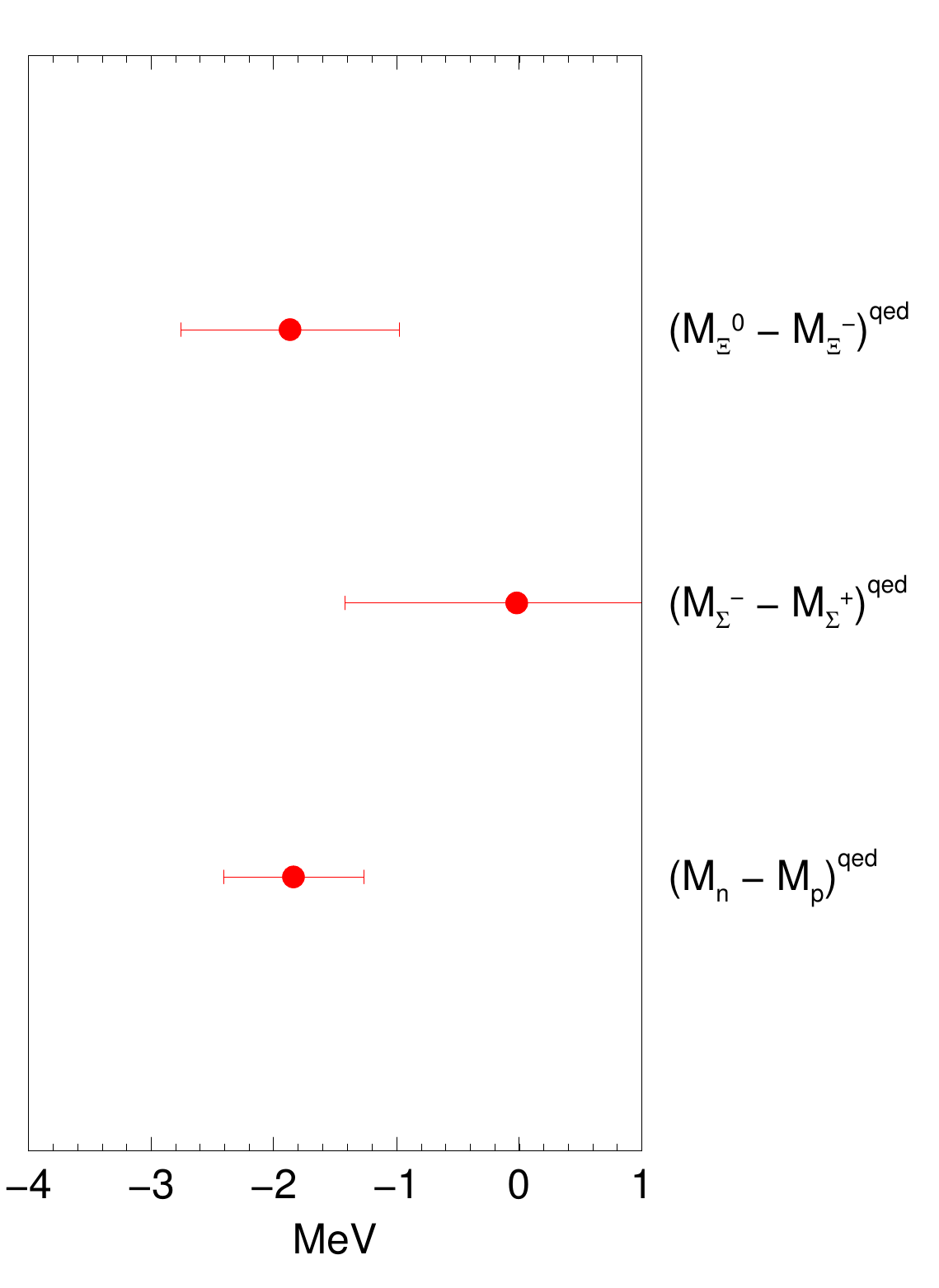}
      \end{center} 

\end{minipage}\hspace*{0.05\textwidth}
\begin{minipage}{0.45\textwidth}

      \vspace*{0.10in}
      \begin{center}
         \includegraphics[width=7.00cm]
            {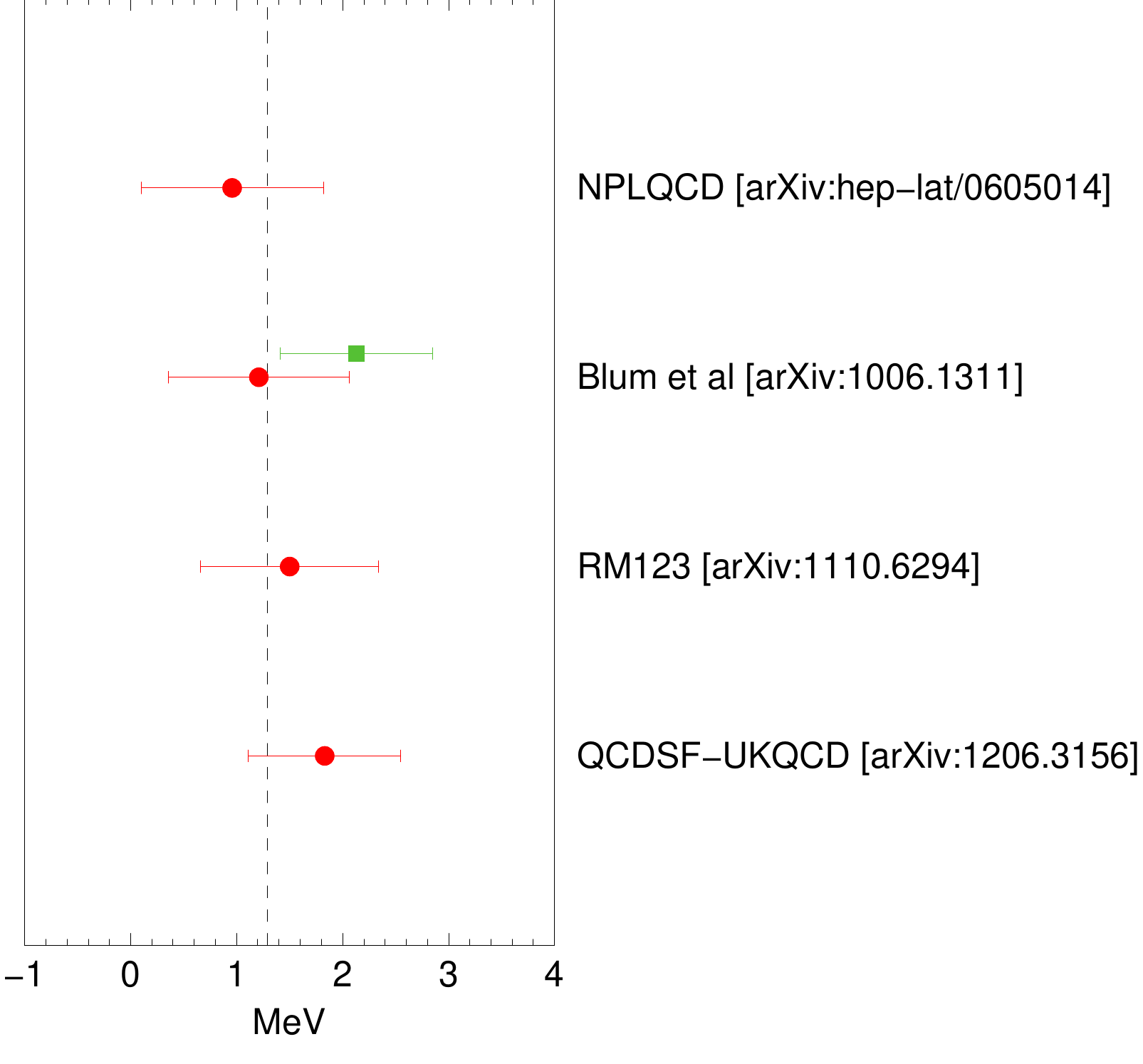}
      \end{center} 

\end{minipage}

\caption{Left panel: QED contribution to the baryon octet mass splittings,
         filled circles. Right panel: comparison of the $n$ -- $p$
         mass difference of the present result (QCDSF--UKQCD or
         bottom number) with NPLQCD, Blum et al., and RM123 
         \protect\cite{beane06a,blum10a,divitiis11a} respectively
         (top to bottom). The filled circles use the QED determination
         of \cite{walker-lourd12a}, while the filled square includes
         the full determination from Blum et al. The vertical dashed
         line is the experimental result.}

\label{results_figs}

\end{figure}

Let us first investigate QED effects. From eq.~(\ref{result_split}),
together with the numerical pseudoscalar meson masses from the previous
paragraph we can determine the `pure' QCD values. Then as we know the
experimental values, \cite{nakamura10a} (as also given on
page~\pageref{pdg_vals}) then from eq.~(\ref{qcd_qed_split})
we can determine the QED contribution to the mass splittings.
This is shown in the left panel of Fig.~\ref{results_figs}.
Thus this indicates that EM effects have the pattern
\begin{eqnarray}
   n(ddu)-p(uud) \approx \Xi^0(ssd) - \Xi^-(ssu) < 0\,, \quad 
   \Sigma^-(dds)-\Sigma^+(uus) \approx 0\,.
\end{eqnarray}

Alternatively \cite{walker-lourd12a} gives a determination
of electromagnetic effects of $n$ -- $p$ of $-1.30(47)\,\mbox{MeV}$
(to be compared with $-1.84(57)\,\mbox{MeV}$ here). In the right panel
of Fig.~\ref{results_figs} we compare our $n$ -- $p$ mass difference
including this determination of the QED contribution, $(M_n - M_p)^{* + \qed}$,
bottom result with the results of  \cite{beane06a,blum10a,divitiis11a}
(top to bottom). The filled square includes the full determination
from that reference. Despite the fact that QED effects are treated
slightly differently in each work good agreement amongst the various
determinations and with the experimental result is found.


\section{Conclusions}


We have introduced a method here to determine `pure' QCD isospin effects in 
\begin{eqnarray}
            n - p\,, \quad \Sigma^- - \Sigma^+\,, \quad \Xi^- - \Xi^0 \,,
\end{eqnarray}
due to differences in $u$ -- $d$ quark masses. This method involves
developing a $SU(3)$ flavour symmetry breaking expansion keeping
the average quark mass $\overline{m}$ constant.
Advantages include the ability to use $2+1$ simulations, i.e.\
$m_u = m_d = m_l$ and use of computationally cheap PQ results.
This expansion appears to be highly convergent, giving encouraging
first results. Clearly the largest errors are due to unknown
QED effects. For more details and numerical results see \cite{horsley12a}.


\section*{Acknowledgements}


The numerical configuration generation was performed using the
BQCD lattice QCD program, \cite{nakamura10b}, on the IBM
BlueGeneL at EPCC (Edinburgh, UK), the BlueGeneL and P at
NIC (J\"ulich, Germany), the SGI ICE 8200 at
HLRN (Berlin--Hannover, Germany) and the JSCC (Moscow, Russia).
The BlueGene codes were optimised using Bagel.
The Chroma software library,
was used in the data analysis. This investigation has been supported partly
by the DFG under contract SFB/TR 55 (Hadron Physics from Lattice QCD)
and by the EU grants 227431 (Hadron Physics2)
and 238353 (ITN STRONGnet). JN was partially supported by EU grant
228398 (HPC-EUROPA2). JMZ is supported by the Australian Research
Council grant FT100100005. We thank all funding agencies.



\end{document}